\documentclass[reprint,prb,showkeys,superscriptaddress,citeautoscript] {revtex4-2}
\usepackage{graphicx}
\usepackage{tikz}
\usepackage{braket}
\usepackage{hyperref}
\hypersetup{colorlinks=true, urlcolor= blue, citecolor=blue, linkcolor= blue, bookmarks=true, bookmarksopen=false}
\usepackage{amsmath,amssymb}
\usepackage{textcomp}
\usepackage{multirow}
\usepackage{enumitem}
\setlist{noitemsep, leftmargin=*}
\usepackage{times}
\begin{document}
		
		\title{Optoelectronic properties of the CuI, AgI and Janus Cu$_2$BrI, and Ag$_2$BrI monolayers by many-body perturbation theory}
		
		\author{Mohammad Ali Mohebpour}
		\email{b92\_mohebi@hotmail.com}
		\affiliation{Department of Physics, University of Guilan, P. O. Box 41335-1914, Rasht, Iran}
		
		\author{Bohayra Mortazavi}
		\email{bohayra.mortazavi@gmail.com}
		\affiliation{Department of Mathematics and Physics, Leibniz Universit{\"a}t Hannover, Appelstra$\beta$e 11,30167 Hannover, Germany}
		
		\author{Xiaoying Zhuang}
		\affiliation{Department of Mathematics and Physics, Leibniz Universit{\"a}t Hannover, Appelstra$\beta$e 11,30167 Hannover, Germany}
		
		\author{Meysam Bagheri Tagani}
		\email{m\_bagheri@guilan.ac.ir}
		\affiliation{Department of Physics, University of Guilan, P. O. Box 41335-1914, Rasht, Iran}

%%%%%%%%%%%%%%%%%%%%%%%%%%%%%%%%%%%%%%%%%%%%%%%%%%%%%%%%%%%%%%%%%%%%%%%%%%%%%%%%%%%%
\begin{abstract}
In an outstanding experimental advance in the field of two-dimensional nanomaterials, cuprous iodide (CuI) and silver iodide (AgI) monolayers have been grown via a novel graphene encapsulation synthesis approach [Adv.Mater.2022, 34, 2106922]. Inspired by this  accomplishment, we  conduct first-principles calculations to investigate the elastic, phononic thermal transport, electronic, and optical properties of the native CuI and AgI and Janus Cu$_2$BrI and Ag$_2$BrI monolayers. Electronic and excitonic optical properties are elaborately studied using the many-body perturbation theory on the basis of GW approximation. Our results indicate that these novel  systems are stable but with soft elastic modulus and ultralow lattice thermal conductivity. It is also shown that the studied monolayers are wide-gap semiconductors with exciton binding energies close to 1 eV. The effects of mechanical straining and electric field on the resulting electronic and optical properties are also analyzed. The presented first-principles results provide a deep understanding of the stability, phononic transport and tunable optoelectronic properties of the native CuI and AgI and Janus Cu$_2$BrI and Ag$_2$BrI monolayers, which can serve as a guide for the oncoming studies.  
\end{abstract}
\maketitle
%%%%%%%%%%%%%%%%%%%%%%%%%%%%%%%%%%%%%%%%%%%%%%%%%%%%%%%%%%%%%%%%%%%%%%%%%%%%%%%%%%%%
	
\section{Introduction}
\par Two-dimensional (2D) materials have attracted the interest of researchers in recent years, for developing next-generation optoelectronic and energy storage/conversion devices due to their outstanding physical properties. Graphene, the most famous 2D material, has been proven to strongly interact with light from microwave to ultraviolet wavelengths, exhibiting wide applications in optoelectronic devices \cite{ bao2012graphene}. To improve the efficiency of optoelectronic devices, novel layered 2D materials have been experimentally synthesized such as transition metal dichalcogenide (TMD) monolayers (like MoS$_2$ \cite{ zeng2012valley}, WS$_2$ \cite{gutierrez2013extraordinary}, and WSe$_2$ \cite{ fang2012high}), metal oxides \cite{zhang2021hexagonal, liu2016metal}, metal perovskite \cite{ ji2019freestanding}, and MXenes \cite{gogotsi2019rise, naguib2021ten}. All aforementioned monolayers are exfoliated from their bulk lattices due to the layered structures, and they usually exhibit unconventional physical properties, distinct from their bulk counterparts.
\par The majority of 2D nanomaterials fabricated to date have layered structures, in which there exist weak van der Waals interlayer interaction. These weak interactions facilitate the exfoliation process and enable the possibility to derive single or few layers. The synthesis of non-layered 2D materials, despite many challenges, not only significantly expands the field of 2D materials, but also opens new horizons in materials science. One of the first synthesized monoelemental non-layered 2D materials was boron allotropes, borophene \cite{mannix2015synthesis, mannix2018borophene}, which revolutionized the study of rechargeable batteries \cite{ zhang2016borophene} and hydrogen storage \cite{joseph2020borophene}. The non-layered monochalcogenides of ZnS \cite{ sun2012fabrication} and PbS \cite{ acharya2013bottom} were also synthesized by bottom-up approach, exhibiting unique features for applications in solar water splitting and field-effect transistor, respectively.  
\par Cuprous iodide, CuI, as a group 11 transition metal halide is a non-layered material and crystallizes in different phases \cite{miyake1952phase}. At temperature below 370 $^0$C
, it is crystallized in cubic $\gamma$-phase. In the temperature range of 370-407 $^0$C
, it is in hexagonal $\beta$-phase and a transition to cubic $\alpha$-phase is happened for higher temperatures.  The bulk $\gamma$-phase of CuI is known for a high thermopower coefficient and substantial optoeletric properties\cite{grundmann2013cuprous}. Very recently, ultrathin nanosheets of $\gamma$-phase CuI has been synthesized on SiO$_2$/Si substrate using a facile physical vapor deposition process \cite{ yao2018synthesis}. It was also shown that the CuI nanosheets can be synthesized on the 2D substrates like WS$_2$ and WSe$_2$. Mustonen \mbox{et al.} \cite{mustonen2022} successfully synthesized monolayer CuI at ambient conditions by a single-step wet-chemical process between graphene layers. However, the mentioned articles did not provide any information on the electronic and optical properties of CuI. Indeed, no optical analysis was performed to determine the value of band gap and its type. In addition, the effect of dimensional reduction on the change of electronic properties of the monolayer in comparison with its bulk counterpart was not discussed.
\par Motivated by the recent advance on the synthesis of monolayer CuI via a novel graphene encapsulation synthesis approach, we herein investigate the electronic, excitonic, optical, and thermal properties of monolayer group 11 transition metal halides, including native CuI and AgI and Janus counterparts of Cu$_2$IBr and Ag$_2$IBr using first-principles calculations. The electronic and optical properties of the considered monolayers are studied using  many-body perturbation calculations, and the excitonic optical properties are calculated by solving the Bethe-Salpeter equation. The dynamical and thermal stabilities of the monolayers are found to be validated by phonon dispersion spectra and ab-initio molecular dynamics simulations, respectively. The quasiparticle band gaps are obtained to be 4.42, 4.26, 3.93, and 3.96 eV for CuI, AgI, Cu$_2$BrI, and Ag$_2$BrI monolayers, respectively. We also examine the effects of strain and external electric field on the spin-orbit induced band splitting and find an increase of more than 100$\%$ in the magnitude of band splitting for the conduction band of Ag$_2$IBr under a uniaxial strain of 3$\%$. In addition, the electric field produces a noticeable band splitting in the non-Janus monolayers. The first bright excitons of CuI, AgI, Cu2BrI, and Ag2BrI monolayers are found to be tightly bound with binding energies of 0.88, 0.92, 0.95, and 0.99 eV, respectively, meaning the super stability of the excitonic states against thermal decomposition at 300 K. To obtain the phonon thermal conductivity of the monolayers, the moment tensor potentials are trained over ab-initio molecular dynamics datasets. Then, the lattice thermal transport is evaluated using the full-iterative solution of the Boltzmann transport equation. The room-temperature thermal conductivity  of CuI, AgI, Cu$_2$BrI, and Ag$_2$BrI monolayers are predicted to be remarkably low,  3.75, 2.27, 3.13, and 1.26 W/mK, respectively. The presented first-principles results provide important understanding about the key physical properties of the non-Janus CuI and AgI and Janus Cu$_2$BrI and Ag$_2$BrI monolayers, which can be useful for their practical application and can  guide oncoming theoretical and experimental works.

\section {Computational methods}
The density functional theory (DFT) calculations are performed using the Vienna Ab-initio Simulation Package (VASP) \cite{kresse1996} on the basis of the generalized gradient approximation (GGA) \cite{perdew1996} and the Perdew-Burke-Ernzerhof functional for solids (PBEsol) \cite{csonka2009}. The plane-wave and \mbox{self-consistent} loop energy cutoffs are set to \mbox{600 eV} and \mbox{10$^{-7}$ eV}, respectively. The optimized lattice structures are obtained using the conjugate gradient algorithm until the Hellman-Feynman forces drop below \mbox{10$^{-3}$ eV/\AA}, considering a fixed vacuum space of 20~\AA~along the thickness of the monolayers. The first Brillouin zone is sampled with a 15$\times$15$\times$1 Monkhorst-Pack k-point mesh \cite{monkhorst1976} based on the convergence of the total energy within \mbox{10$^{-6}$~eV}.
The electronic band structure is also evaluated using the Heyd-Scuseria-Ernzerhof (HSE06) hybrid functional with the default mixing parameter \mbox{($\alpha=0.25$)} \cite{krukau2006in}. The charge transfer between atoms is determined using the Bader analysis \cite{henkelman2006}.

The first-principles many-body perturbative calculations are carried out using the single-shot GW method \cite{shishkin2007}, referred to as G$_0$W$_0$. In these calculations, the quasiparticle (QP) band gap should be carefully converged with respect to the number of virtual bands, number of frequency grid points, kinetic energy cutoff, and k-point mesh. After the convergence test, a set of 168, 204, 152, and 186 virtual bands is employed in the G$_0$W$_0$ calculations for CuI, AgI, Cu$_2$BrI, and Ag$_2$BrI monolayers, respectively. The number of frequency grid points is selected to be 96. The kinetic energy cutoff for the plane wave and the response function is set to be 600 and \mbox{200 eV}, respectively. The Brillouin zone is integrated with a 9$\times$9$\times$1 $\Gamma$-centered k-point mesh. The QP band structure of the G$_0$W$_0$ calculations is interpolated using the maximally localized Wannier functions (MLWFs), as implemented in the WANNIER90 code \cite{mostofi2014}. Here, the number of Wannier bands is selected to be 24, and the $sp^3d^2$ hybrid orbitals are chosen for the initial projections. The excitonic optical properties are studied by calculating the macroscopic dielectric function, given as \mbox{$\varepsilon(\omega)=\varepsilon_1 (\omega)+i\varepsilon_2 (\omega)$}, through solving the Bethe-Salpeter equation (BSE) \cite{rohlfing2000} on top of the G$_0$W$_0$ eigenvalues. Here, the Tamm-Dancoff approximation (TDA) is used, which excludes the resonant-antiresonant coupling \cite{onida2002}. The imaginary part of the dielectric function is obtained by summation over the empty conduction bands, and the real part is transformed by the Kramers-Kronig relation, as elaborately discussed in our previous study \cite{mohebpour2022tran}. To check the effects of \mbox{self-consistency} in the excitonic properties, we perform the fully self-consistent GW (i.e. QPGW). At this level, the QP energies and one-electron orbitals are updated 4 times in the calculations of the Green’s function ($G$) and screened Coulomb interaction ($W$). The 15 highest occupied valence bands and the 15 lowest unoccupied conduction bands are taken into account to get a converged BSE spectrum. For optical calculations, the spin-orbit coupling (SOC) is considered.

To evaluate the phononic properties, the moment tensor potentials (MTPs) \cite{shapeev2016moment} are developed using the MLIP package \cite{novikov2020mlip}. The ab-initio molecular dynamics (AIMD) simulations are carried out with a time step of 1 fs and a \mbox{Monkhorst-Pack} k-point mesh of 2$\times$2$\times$1 over 4$\times$3$\times$1 supercells. Two separate AIMD calculations are conducted with the controlled temperature from 2 to 100 K and from 300 to 1000 K, each for 1000 time steps. Complete AIMD trajectories are then subsampled, and around 800 configurations are selected to train the MTPs. The phonon dispersion relations and harmonic 2$^{nd}$ order interatomic force constants are obtained using the PHONOPY code \cite{togo2015first} over 6$\times$6$\times$1 supercells \cite{mortazavi2020ex}.
The anharmonic 3$^{rd}$ order interatomic force constants are calculated using the MTPs over 6$\times$6$\times$1 supercells \cite{mortazavi2021ac}, considering the interactions with eleventh nearest neighbors. The lattice thermal transport is evaluated using the \mbox{full-iterative} solution of the Boltzmann transport equation (BTE) via employing the ShengBTE package \cite{li2014shengbte}, considering the isotope scattering.

\section{Results and discussion}
\subsection{Structural properties}

\begin{figure}
	\centering{
	\includegraphics[width=0.46\textwidth]{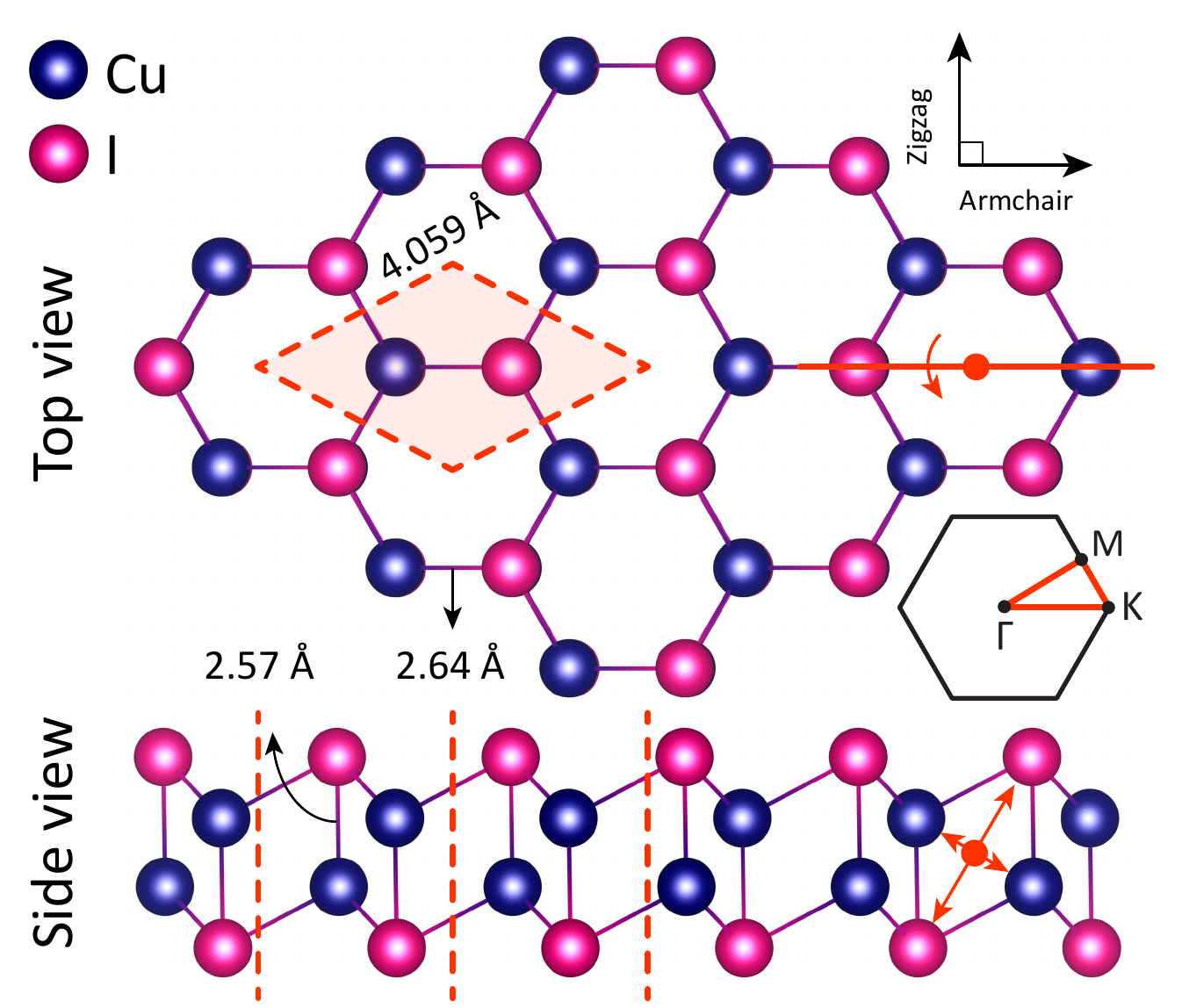}}
	\caption{Top and side views of the relaxed CuI monolayer with the first Brillouin zone. The orange dashed lines show the unit cell. The orange arrows show the mirror and inversion symmetries. The \mbox{in-plane} and \mbox{out-of-plane} Cu-I bond lengths are given.}
	\label{struct}
\end{figure}

After the experimental achievement by Mustonen and coworkers \cite{mustonen2022}, we found CuI and AgI monolayers very interesting for exploring their physical properties. These systems belong to the group 11 transition-metal halide MX
\mbox{(M = Cu, Ag, Au; X = Cl, Br, I)} 2D lattices, among which MCl and AuX monolayers are thermally and dynamically unstable \cite{huang2020g}. We also constructed two Janus monolayers based on CuI and AgI native monolayers to seek fascinating properties caused by inversion symmetry breaking. For this purpose, we replaced the I atoms on one side of the native monolayers with Br atoms, because with Cl atoms the stability is expected to be low. By and large, in this work, we study \mbox{non-Janus} CuI and AgI and Janus Cu$_2$BrI and Ag$_2$BrI monolayers, which all have highly symmetrical hexagonal lattice. Fig.~\ref{struct} illustrates the top and side views of the \mbox{fully-relaxed} CuI monolayer as a representative.
%Similar hexagonal lattice was previously observed in the synthesized AlSb honeycomb structure \cite{qin2021re}.
Obviously, the non-Janus monolayers show inversion and mirror symmetries, however, the Janus ones have only mirror symmetry. That is to say, the inversion symmetry is broken in the Janus monolayers due to the replacement of the I atom by Br atom. The relaxed lattice constant of CuI, AgI, Cu$_2$BrI, and Ag$_2$BrI monolayers according to the PBEsol functional is predicted to be 4.059, 4.402, 3.962, and \mbox{4.326 \AA}, respectively. Based on experimental tests \cite{mustonen2022}, the distances between the Cu atomic planes and the Cu-I bond lengths along the \mbox{in-plane} and \mbox{out-of-plane} directions are 1.63$\pm$0.35, 2.67$\pm$0.16, and \mbox{2.55$\pm$0.49 \AA}, respectively, which are in good agreement with the corresponding values of 1.345, 2.642 and \mbox{2.566 \AA}, respectively.

We firstly study the stability of CuI, AgI, Cu$_2$BrI, and Ag$_2$BrI monolayers on the basis of the phonon dispersion relations. The obtained phonon dispersion spectra along the high symmetry points of the first Brillouin zone are depicted in Fig.~S1 of the supporting information document. Apparently, there is no imaginary frequency in the phonon spectra, confirming their dynamical stability. The thermal stability of monolayers is then examined by the AIMD simulations conducted at \mbox{500 K}. As shown in Fig.~S2, during the simulations, the total energy of the monolayers fluctuates around the mean values, confirming their thermal stability up to \mbox{500 K}. Interestingly, the $C_{11}$ ($C_{12}$) elastic constant of CuI, AgI, Cu$_2$BrI, and Ag$_2$BrI monolayers is calculated to be 47~(21), 26~(15), 44~(21), and \mbox{24~(15) N/m}, which reveal that these novel 2D systems are soft and include weak bonding interactions. Moreover, it is found that the systems with Cu atoms show distinctly higher rigidity than the counterparts based on Ag atoms. According to the aforementioned findings, the CuI, AgI, Cu$_2$BrI, and Ag$_2$BrI monolayers are very stable but generally among soft nanomaterials.

\subsection{Electronic and optical properties}

\begin{figure*}
	\centering{
	\includegraphics[width=0.9\textwidth]{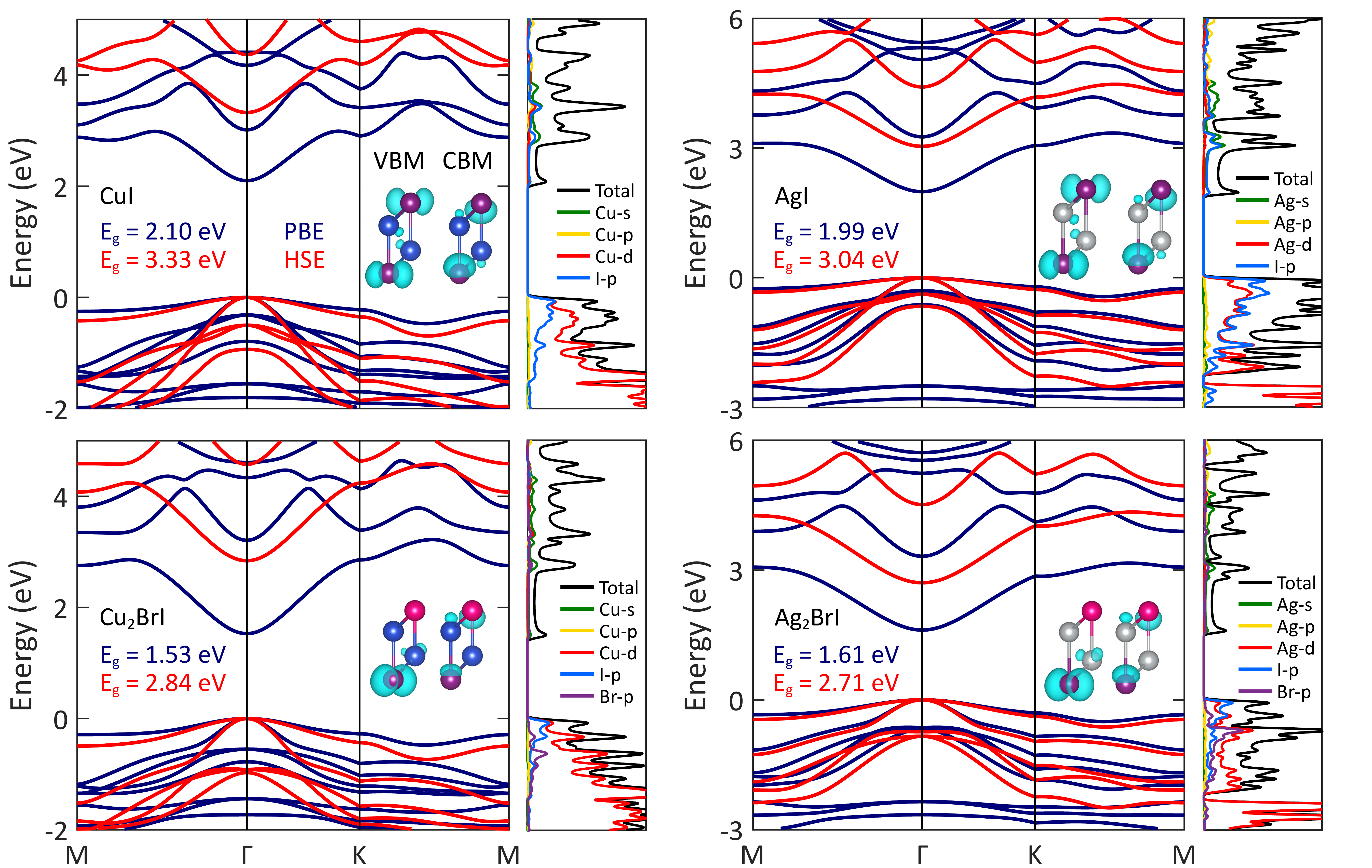}}
	\caption{Electronic band structures of the monolayers at the PBE (blue lines) and HSE06 (red lines) levels together with the density of states at the PBE level. The VBMs are set to zero. The squares of the wave functions at the VBM and CBM are shown in the insets.}
	\label{band}
\end{figure*}

After assuring the stability of the monolayers, we turn our attention to their electronic properties. Fig.~\ref{band} displays the band structures of the monolayers together with their total and partial density of states (DOS). At the PBE (HSE06) level, the monolayers are semiconductors with direct band gaps of
\mbox{2.10 (3.33)}, 1.99 (3.04), 1.53 (2.84), and \mbox{1.61 (2.71) eV} for CuI, AgI, Cu$_2$BrI, and Ag$_2$BrI, respectively. Except for the band gaps, the characteristics of the band structures are the same at both levels. That is to say, including a fraction (0.25) of the exact Hartree-Fock exchange energy in the DFT formulation only corrects the band gaps by 1.23, 1.05, 1.31, and \mbox{1.10 eV} for CuI, AgI, Cu$_2$BrI, and Ag$_2$BrI, respectively. For all the monolayers, the valence band maximum (VBM) and the conduction band minimum (CBM) are located at the $\Gamma$ point. The conduction band edges are quadratically dispersed, that show free electrons. The valence band edges are nearly flat, that represent localized holes. Although the valence bands edges at the $\Gamma$ point show double degeneracy, there is a tiny gap of \mbox{$10^{-3}$ eV} between the heavy and light hole subbands. It is worth noting that the band gaps calculated for CuI and AgI monolayers agree very well with those reported in a previous study \cite{huang2020g}.

From the DOS, it is seen that the CBMs are equally contributed by the $s$-orbitals of the Cu/Ag atoms and the\mbox{ $p$-orbitals} of the Br/I atoms while the VBMs are mostly controlled by the \mbox{$d$-orbitals} of the Cu/Ag atoms and the \mbox{$p$-orbitals} of the I atoms. Therefore, the optical selection rules allow the transitions between the $p$-orbitals of the halogen atoms and the $s$-orbitals of the transition metal ones. No considerable participation from Br atoms is found near the edge of valence band while contributing the most at the edge of conduction band in the Janus monolayers, which is in line with the reduction of the band gap by substituting Br atoms. The shapes of the wave functions at the VBMs and CBMs support the above discussion. As shown in the insets of Fig.~\ref{band}, at the VBMs of the non-Janus monolayers, the wave functions are shaped like dumbbells, which are distributed along the \mbox{x-direction} and centered on the I atoms, showing the \mbox{$p_x$-orbitals} of the I atoms. At the CBMs, the wave functions are formed spherically on the Cu/Ag atoms and hemispherically on the I atoms, referring to as $s$ and $p_z$ orbitals, respectively. For Ag$_2$BrI and Cu$_2$BrI Janus monolayers, the wave functions of the VBMs exhibit a hybridization between the $p_x$ and $p_y$-orbitals of the I atoms. At the CBMs, one can see the predominant contribution of the $p_z$-orbitals of the halogen atoms.

\begin{figure}
	\centering{
	\includegraphics[width=0.45\textwidth]{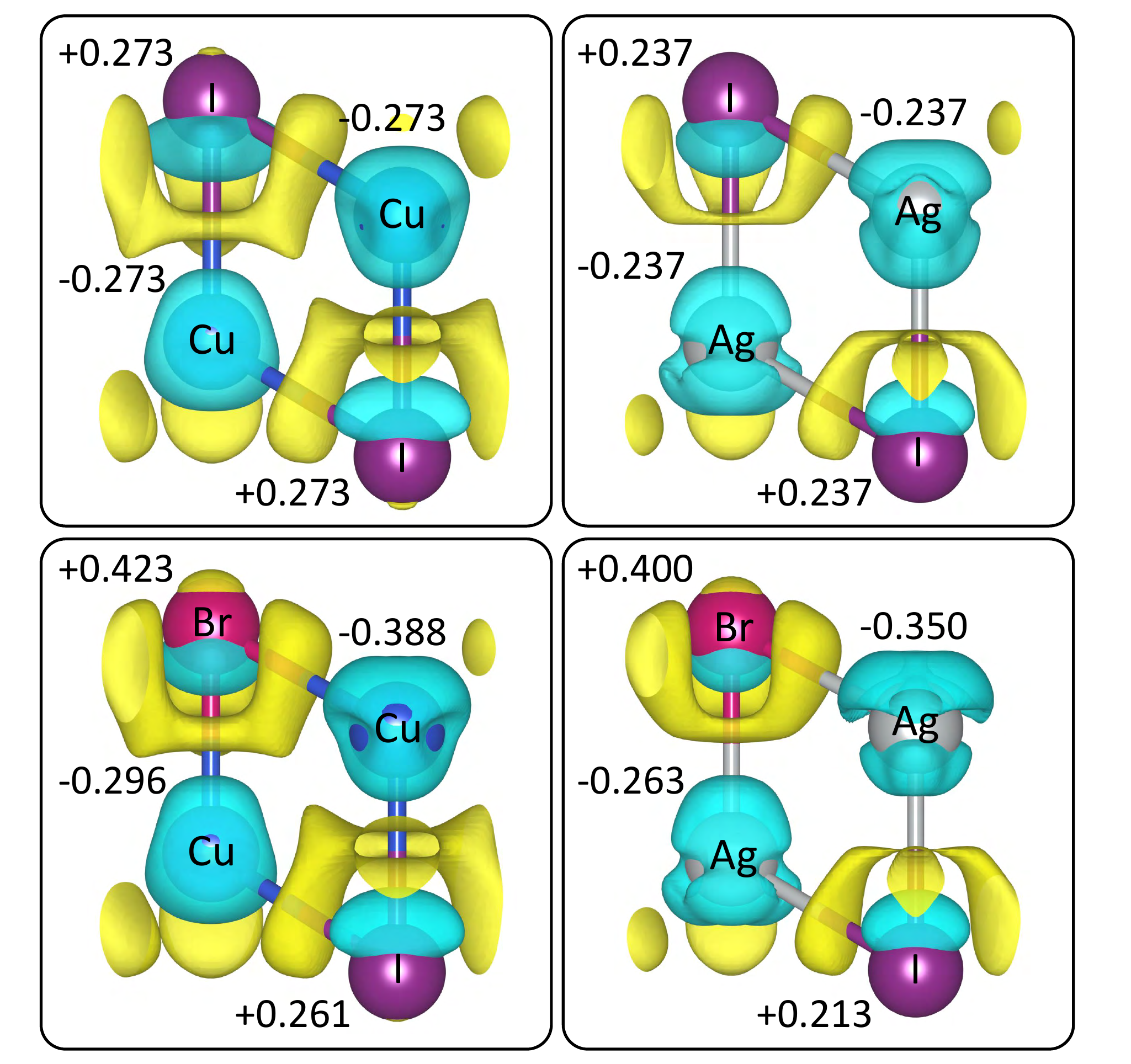}}
	\caption{Charge density difference of the monolayers with the Bader charge transfer between atoms. The isosurface of charge accumulation and depletion was set to 5$\times$10$^{-3}$ e/bohr$^3$.}
	\label{CDD}
\end{figure}

The charge density difference (CDD) isosurface of the monolayers is depicted in Fig.~\ref{CDD}, where the yellow and cyan colors denote the charge accumulation and depletion, respectively. Apparently, the highest charge accumulation is located at the center of the atomic bonds, indicating the covalent bonding characteristics. The highest charge depletion is around the transition metal atoms. In accordance with the CDD analysis, for CuI structure, we find that \mbox{0.273 $e$} is transferred from the Cu to I atoms. While, for AgI structure, \mbox{0.237 $e$} is transported from the Ag to I atoms. These are in line with the larger electronegativity of the I (2.66) atom compared to the Cu (1.90) and Ag (1.93) transition metals. In Cu$_2$BrI Janus monolayer, the charge transfer from the two neighboring Cu atoms to the uppermost and lowermost layers of the halogen atoms is different. That is to say, the Br atoms receive more charge compared to the I atoms due to the larger electronegativity of Br (2.96) atom.
Also, in Ag$_2$BrI Janus monolayer, the two neighboring Ag atoms lose different amounts of charge in such a manner that 65\% of the total transferred charge is delivered to the Br atoms and the rest \mbox{(0.213 $e$)} is sent to the I atoms. These differences result in a potential gradient normal to the basal plane, forming an intrinsic internal electric field between the uppermost and lowermost layers. The intrinsic \mbox{out-of-plane} electric field caused by inversion symmetry breaking in the Janus monolayers might lead to spin-valley splitting and Rashba spin splitting when combined with the SOC.

\begin{figure*}
	\centering{
	\includegraphics[width=0.9\textwidth]{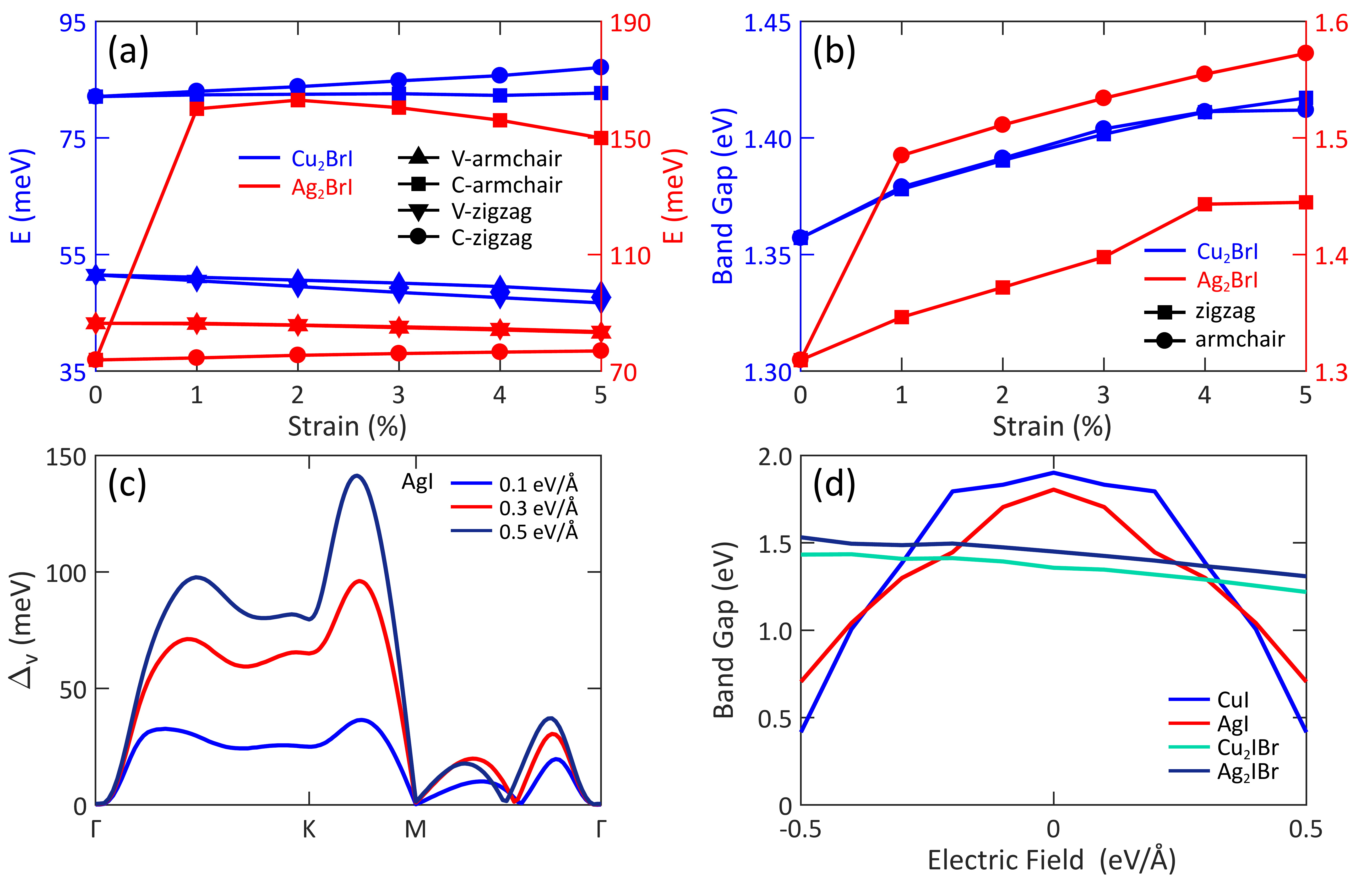}}
	\caption{(a) Spin-orbit induced valence and conduction bands splitting and (b) band gap variation of Cu$_2$BrI and Ag$_2$BrI Janus monolayers as a function of uniaxial strain along the armchair and zigzag directions. (c) Spin-orbit induced valence band splitting of AgI monolayer under different electric fields. The rest of the results are available in Fig.~S5.(d) Band gap variation of the monolayers as a function of electric field.}
	\label{soc}
\end{figure*}

We check out the effects of the SOC in the band structures. By including the SOC at the PBE (HSE06) level, the numbers of energy bands are doubled and the band gaps are reduced to 1.86 (3.03), 1.72 (2.73), 1.31 (2.58), and 1.37 (2.44) eV for CuI, AgI, Cu$_2$BrI, and Ag$_2$BrI monolayers, respectively. As shown in Fig.~S3 (Fig.~S4), for CuI and AgI monolayers, the SOC cannot remove the double degeneracy of the individual energy states while for the Janus monolayers, the bands are split. This spin splitting, referred to as Zeeman-type splitting, is in fact a consequence of the SOC combined with inversion symmetry breaking. At the PBE level, the spin splitting constant at the $\Gamma$ point is zero but at the $K$ point, the spin splitting of valence (conduction) band edge is \mbox{12 (28)} and \mbox{13 (45) meV} for Cu$_2$BrI and Ag$_2$BrI monolayers, respectively.
The SOC also increases the tiny gap observed between the heavy and light hole subbands at the $\Gamma$ point up to 400 and \mbox{450 meV} for the non-Janus and Janus monolayers, respectively. At the HSE06 level, the strength of SOC is predicted to be higher. Hence, the reported values for spin splitting will be slightly larger at this level. By and large, one can conclude that the main effect of the SOC is on the valence bands owing to the hybridization between the $d$-orbitals of the Cu/Ag atoms and the $p$-orbitals of the I atoms. The conduction bands are not influenced by the SOC due to the predominant participation of the $s$-orbitals of the Cu/Ag atoms.

In experimental syntheses, 2D materials are grown on different kinds of metal or insulator substrates. These substrates can affect the intrinsic properties of materials. The most common effect of the substrate is applying strain to the overlayer. If the substrate is active, the interactions between the substrate and 2D material can strongly affect the electronic properties of the overlayer.
In this part, we examine the dependence of spin-orbit induced band splitting on the applied strain. It should be noted that strain is only applied in the form of uniaxial in the armchair and zigzag directions because it can break the inversion symmetry of materials. The combination of spin-orbit with inversion symmetry breaking produces attractive effects such as Rashba \cite{mera2019zeeman} and valleytronic \cite{he2021two}. Inversion symmetry is inherently broken in Janus structures, and as noted, the SOC causes band splitting at the $K$ points. The maximum band splitting in Cu$_2$BrI and Ag$_2$BrI Janus monolayers as a function of strain is plotted in \mbox{Fig.~\ref{soc} (a)}. Irrespective of direction, strain slightly reduces the valence band splitting. In contrast, we see an increase in the conduction band splitting with strain.
The stronger effect of spin-orbit interaction in the conduction band is due to the predominant role of the transition metal in the conduction band, which is observable from the DOS.
The rate of increase for strain in the armchair direction is slightly higher than that in the zigzag direction. In addition, the increase in Ag$_2$BrI structure is much more intense than in Cu$_2$BrI monolayer. A strain of about \mbox{1$-$3\%} in Ag$_2$BrI monolayer can cause an increase of more than 100\% in the magnitude of band splitting for the conduction band. Strong spin-orbit interactions, inherent electric field, and increased inversion symmetry breaking by strain are the main reasons for the observed increase.

We additionally examine the spin-orbit induced band splitting of CuI and AgI monolayers in terms of the applied uniaxial strain. Interestingly, although strain can slightly alter the symmetries of the structures, no band splitting induced by spin-orbit interaction is observed in these monolayers \mbox{(Data not shown)}. The softness of the materials, which was also confirmed in experimental reports \cite{mustonen2022}, is a major obstacle for engineering the band structures and creating the \mbox{Zeeman-type} or \mbox{Rashba-type} spin splitting with strain. Our evaluation shows that applying strain is not a good approach for tuning the properties of CuI and AgI monolayers. Hence, they can be grown on substrates with different lattice constants while showing the same electronic properties.%These novel 2D materials can be used in the flexible optoelectronic industry and can operate under finite strains without losing efficiency.

\mbox{Fig.~\ref{soc} (b)} illustrates the variations of the band gaps of Cu$_2$BrI and Ag$_2$BrI Janus monolayers as a function of strain. Apparently, an increase of 20\% (10\%) is observed in the band gap of Ag$_2$BrI monolayer under the strain of 5\% in the armchair (zigzag) direction. This is attributed to the presence of a heavy transition metal such as silver and the absence of an inversion symmetry center in this Janus monolayer, which are critical for the noticeable change in the band gap. The variation of the band gap in Ag$_2$BrI monolayer is much more intense than that in Cu$_2$BrI, which is in line with the smaller elastic constants of Ag$_2$BrI monolayer. In other structures, the changes in band gaps under strain up to 5\% are less than 3\% \mbox{(Data not shown)}.

We also explore the spin-orbit induced band splitting in the presence of an external electric field. In contrast to strain, the electric field produces a noticeable band splitting in CuI and AgI monolayers. The splitting is visible throughout the Brillouin zone, as shown in \mbox{Fig.~\ref{soc} (c)} and Fig.~S5. For \mbox{AgI (CuI)} monolayer, the largest valence band splitting occurs in the $K-M$ path and can reach \mbox{140 (73) meV} under an electric field of \mbox{0.5 eV/\AA}, which is comparable to transition metal dichalcogenide monolayers. The largest conduction band splitting occurs at the $K$ point. Also, a \mbox{Rashba-type} band splitting is observed in the $\Gamma-K$ path under strong electric fields. Comparing the band splitting in CuI and AgI monolayers shows that it is directly related to the transition metal atoms. The larger the atomic number is, the greater the band splitting will be. However, in the Janus structures, the dependence of band splitting on the electric field is very weak as shown in Fig.~S5. Due to the lack of an inversion center, there is an inherent electric field in these structures that leads to a noticeable band splitting as discussed above. Therefore, the external electric field cannot significantly change the band splitting. The interesting point is that, unlike the non-Janus structures, increasing the strength of the external field has no effect on the magnitude of band splitting. In addition, the direction of the electric field does not have a significant effect on the results.

The variation of the band gaps as a function of external electric field is shown in \mbox{Fig.~\ref{soc} (d)}. In CuI and AgI \mbox{non-Janus} monolayers, due to the inversion symmetry and the arrangement of atoms in the lower and upper planes, the changes in the band gaps are independent of the direction of electric field. By increasing the intensity of electric field, the band gaps decrease and the reduction is more for CuI monolayer. On the contrary, in the Janus structures, the changes in the band gaps in addition to the magnitude of electric field depend on its direction. A linear increase is observed in the band gaps with the electric field in the \mbox{-z direction}. While we see a decrease in the band gaps by increasing the electric field in the \mbox{+z direction}.
The band gaps reduction directly depends on the difference in charge transferred between two halogen atoms. As the electric field increases in the \mbox{+z direction}, we observe more charge transfer towards the Br atoms, which reduces the band gaps. \mbox{In contrast}, the external electric field in the \mbox{-z direction} is in the opposite direction of the intrinsic electric field caused by the dipole inside the monolayers, which reduces the difference in the charge accumulated on halogen atoms, giving rise to the band gaps.

To capture the role of electron-electron exchange interaction in the electronic structure, we step beyond the PBE and HSE06 levels of theory using the \mbox{non-self-consistent} GW approach, known as G$_0$W$_0$. The G$_0$W$_0$ band structures of the monolayers in the absence of the SOC interaction are available in Fig.~S6. As can be noticed, the general shapes of the band structures are similar to those calculated with the PBE functional provided in Fig.~\ref{band}. All the monolayers still have direct band gaps at the $\Gamma$ point. However, the values of band gaps are different from the PBE calculations due to the self-energy corrections. At this level, the QP band gaps are obtained to be 4.42, 4.26, 3.93, and \mbox{3.96 eV} for CuI, AgI, Cu$_2$BrI, and Ag$_2$BrI monolayers, respectively. Such large self-energy corrections of 2.32, 2.27, 2.40, and \mbox{2.35 eV} emphasize the importance of the electron-electron interaction in these structures. Also, it is found that increasing the atomic weight of the halogen (transition metal) atom increases (decreases) the QP band gap of the monolayer. Owing to the slight deformation in the bands dispersion, a rigid shift of the conduction bands with respect to the VBMs does not describe the self-energy effects properly. Including the SOC reduces the QP band gaps down to 4.02, 3.85, 3.59, and 3.61 eV for CuI, AgI, Cu$_2$BrI, and Ag$_2$BrI monolayers, respectively. As expected, the structure with heavier atoms has the strongest SOC. To better compare, we listed the band gaps calculated at different levels in Table~\ref{gaps}.

\begin{table*}
	\normalsize
	\caption{Band gaps calculated for the considered monolayers at different levels of theory in the unit of eV.}
	\begin{tabular*}{\textwidth}{@{\extracolsep{\fill}}lllllllll}
	\hline
	Structure&PBE&PBE$+$SOC&HSE06&HSE06$+$SOC&G$_0$W$_0$&G$_0$W$_0+$SOC&Optical \\
	\hline\hline
	
	CuI&	    2.10&	1.86&	3.33&	3.03&	4.42&	4.02&   3.14\\ 
	AgI&	    1.99&	1.72&	3.04&	2.73&	4.26&	3.85&   2.93\\ 
	Cu$_2$BrI&	1.53&	1.31&	2.84&	2.58&	3.93&	3.59&   2.64\\ 
	Ag$_2$BrI&	1.61&	1.37&	2.71&	2.44&	3.96&	3.61&   2.62\\ \hline
	\end{tabular*}
	\label{gaps}
\end{table*}

\begin{figure*}
	\centering{
	\includegraphics[width=0.98\textwidth]{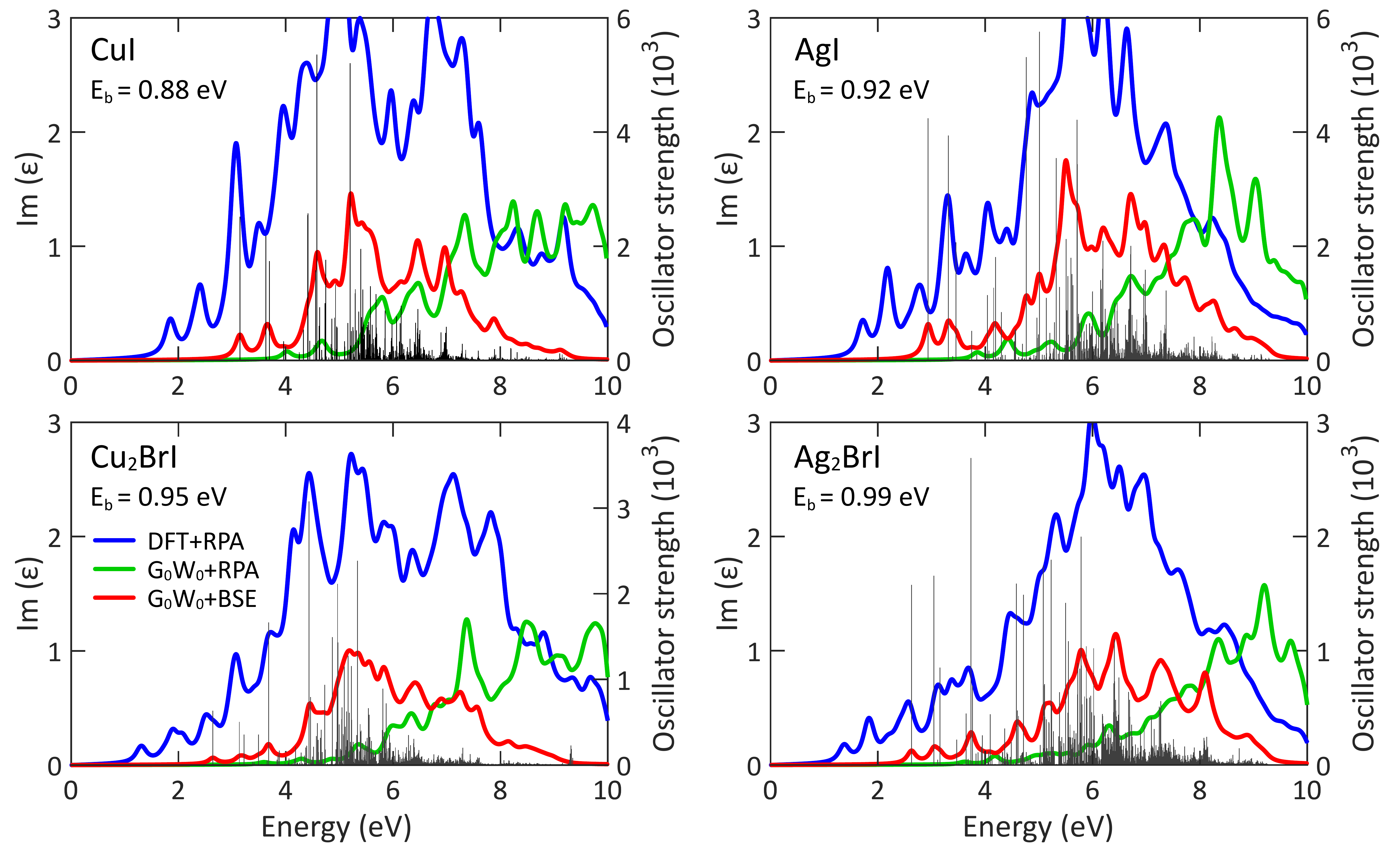}}
	\caption{Imaginary part of the dielectric function of the monolayers at different levels of theory including the DFT$+$RPA (electron$-$electron and electron$-$hole interactions excluded), G$_0$W$_0+$RPA (electron$-$electron interaction included and electron$-$hole interaction excluded), and G$_0$W$_0+$BSE (electron$-$electron and electron$-$hole interactions included) together with the optical oscillator strength. The exciton binding energy ($E_b$), the difference between the G$_0$W$_0+$SOC band gap and the optical gap, is given for each monolayer.}
	\label{IM}
\end{figure*}

\begin{figure}
	\centering{
	\includegraphics[width=0.48\textwidth]{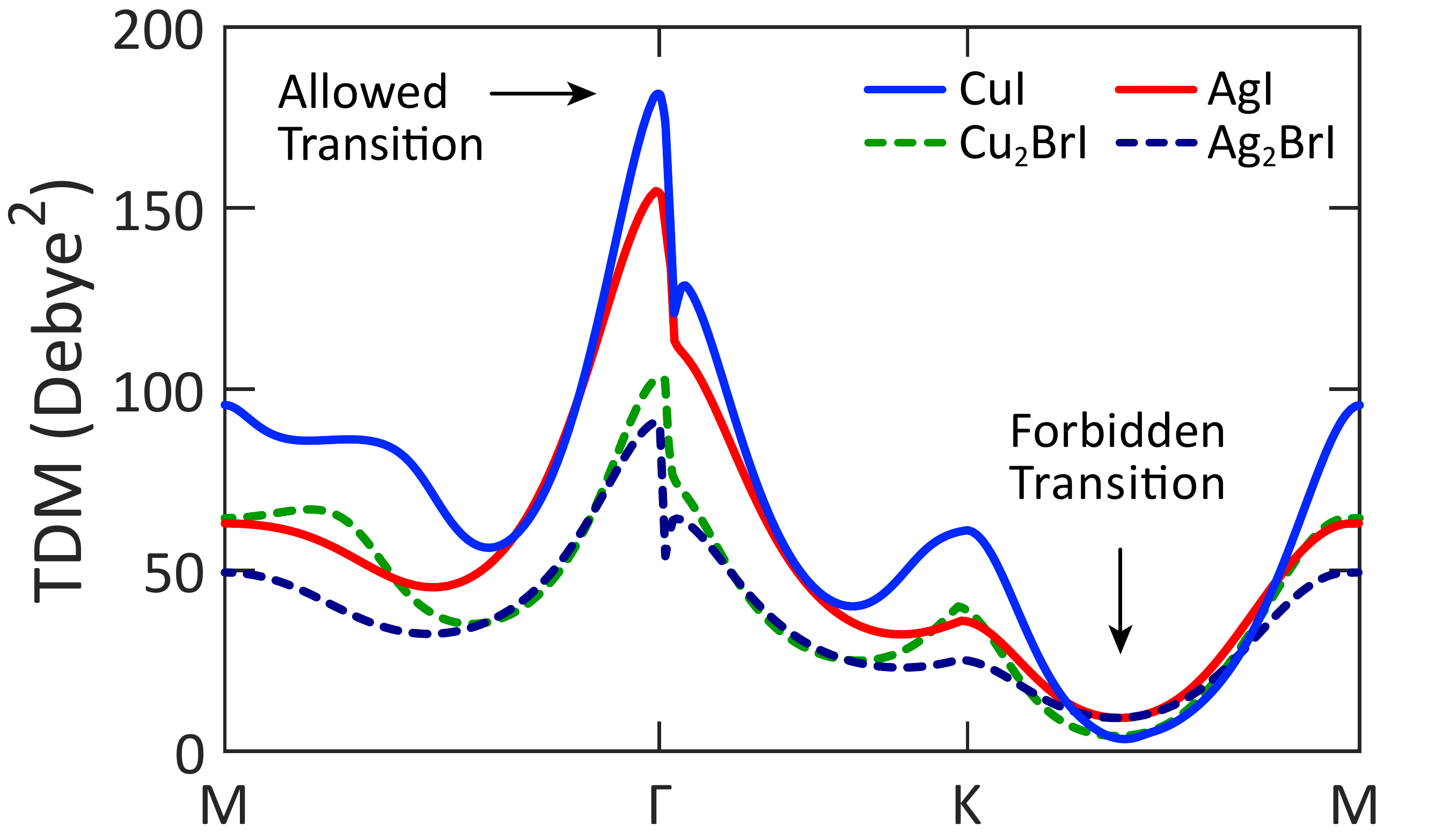}}
	\caption{Amplitude of transition dipole moment of the monolayers along the high symmetry points of the Brillouin zone.}
	\label{tdm}
\end{figure}

We study the excitonic optical properties of the monolayers by solving the Bethe-Salpeter equation over the G$_0$W$_0$ eigenvalues (i.e., G$_0$W$_0+$BSE), which includes the electron-electron and electron-hole interactions. These interactions have been shown to play a vital role in the optical properties of 2D materials \cite{min2019tunable, xu2017elec}. To clarify the role of the many-body effects, we also calculated the optical spectra of the monolayers using the random-phase approximation (RPA) over the G$_0$W$_0$ \mbox{(i.e., G$_0$W$_0+$RPA)} and the DFT \mbox{(i.e., DFT$+$RPA)}, where the aforementioned interactions are excluded. Owing to the symmetric and isotropic crystal structure of the monolayers, the optical coefficients along the x-direction \mbox{($E\parallel x$)} are identical to those along the \mbox{y-direction} \mbox{($E\parallel y$)}.
On the other hand, due to the strong depolarization effect in the low-dimensional systems for light polarization perpendicular to the plane \mbox{($E\parallel z$)}, the optical coefficients along the \mbox{z-direction} are negligible. Therefore, we only study the optical response of the monolayers for the \mbox{in-plane} \mbox{x-polarized} light. Fig.~\ref{IM} illustrates the imaginary part of the dielectric function of the monolayers at three different levels of theory for light polarized along the \mbox{x-direction}. As it is obvious, at the DFT$+$RPA level, the optical spectra of the monolayers show numerous peaks, which are associated with the doubling of bands in the presence of the SOC interaction. At this level, the first peak appears at 1.85, 1.71, 1.30, and \mbox{1.36 eV} for CuI, AgI, Cu$_2$BrI, and Ag$_2$BrI monolayers, respectively. These peaks correspond to direct transitions from the highest valence band to the lowest conduction band at the $\Gamma$ point. Including the \mbox{electron-electron} interaction \mbox{(i.e., G$_0$W$_0+$RPA)} results in a blue shift in the optical spectra and reduces the intensity of the peaks overestimated by the DFT$+$RPA level. The \mbox{electron-electron} interaction also decreases the number of peaks.
On the other hand, considering the \mbox{electron-hole} interaction leads to a cancelation effect and subsequently a redshift in the optical spectra. The \mbox{electron-hole} interaction modifies the general shape of the spectra. At this level (\mbox{i.e., G$_0$W$_0+$BSE}), the first peaks, referred to as optical gaps, appear at 3.14, 2.93, 2.64, and \mbox{2.62 eV}, which correspond to tightly bound excitons with binding energies of 0.88, 0.92, 0.95, and \mbox{0.99 eV} for CuI, AgI, Cu$_2$BrI, and Ag$_2$BrI monolayers, respectively.
This is very interesting because in 2D materials the binding energy of the first bright exciton is directly proportional to the band gap \cite{long2021e,jiang2017}. Basically, the smaller the QP band gap is, the stronger the Coulomb screening becomes. Stronger Coulomb screening naturally results in a smaller exciton binding energy and a larger exciton Bohr radius. However, our results show that the substitution of the Br atoms reduces the QP band gap of the monolayer and increases the exciton binding energy. This is probably because the intrinsic electric field caused by inversion symmetry breaking reduces the Coulomb screening, leading to a larger exciton binding energy. The calculated binding energy for each monolayer indicates the first bright exciton to be strongly confined in an area with a Bohr radius smaller than the bond length, referring to as \mbox{Frenkel-type} exciton, meaning the super stability of the excitonic states against thermal decomposition at \mbox{300 K}. The obtained optical gaps reveal the potential application of the monolayers in the ultraviolet region.

The optical oscillator strength of the monolayers is also depicted in Fig.~\ref{IM}. In spectroscopy, oscillator strength is a dimensionless quantity that expresses the probability of absorption or emission of electromagnetic radiation in transitions between the energy levels. Hence, direct (bright) transitions will have large oscillator strengths. As can be seen, there are many excitonic states with large oscillator strengths. For instance, the oscillator strength of the first peak is 2519, 4243, 634, and 1578 for CuI, AgI, Cu$_2$BrI, and Ag$_2$BrI monolayers, respectively. This means that the first peak of the optical spectra corresponds to a bright exciton. On the contrary, we have many dark transitions with small oscillator strength, which are known as spin-forbidden dark excitons. Among the most important ones, one can refer to the spin-forbidden dark exciton below the optical gap at 3.13, 2.92, 2.63, and 2.61 eV for CuI, AgI, Cu$_2$BrI, and Ag$_2$BrI monolayers, respectively. These excitons have lower energy
\mbox{($\sim0.01$ eV)} than the first bright excitons. Therefore, the ground state excitons of these monolayers are dark.

Fig.~\ref{tdm} provides the amplitude of transition dipole moment (TDM) of the monolayers from the highest occupied valence band to the lowest unoccupied conduction band. The TDM is a complex vector quantity that includes the phase factors associated with the two states, and the amplitude of TDM gives the probability of transition between the two states. Obviously, for all the monolayers, the amplitude of TDM is very small in the $K\rightarrow M$ path, implying no optical absorption between the two states in this path. Meanwhile, the largest amplitude is located at the $\Gamma$ point, revealing the high probability of transition between the VBM and CBM. The transition at the $\Gamma$ point is known as allowed transition.

\begin{figure}
	\centering{
		\includegraphics[width=0.48\textwidth]{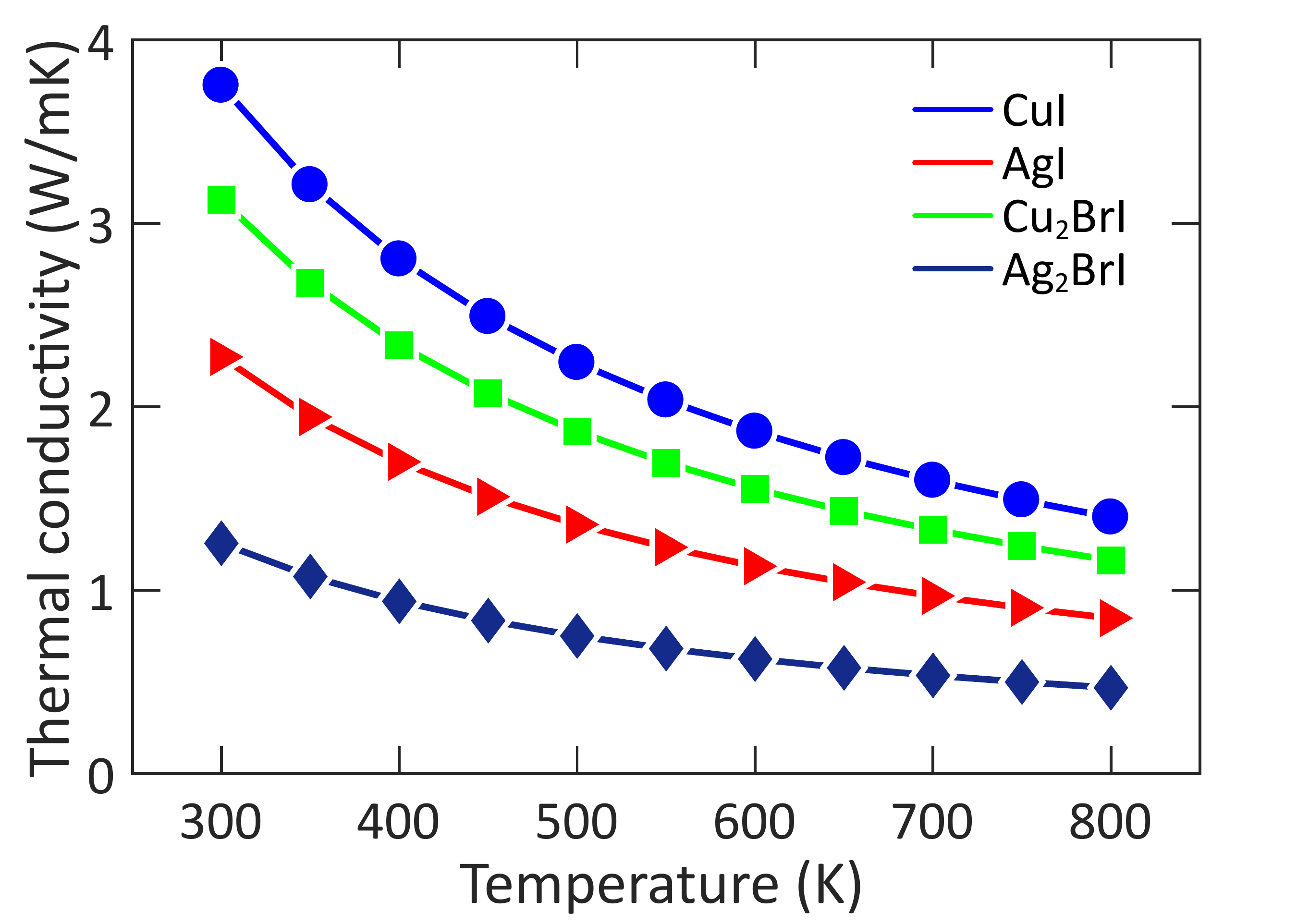}}
	\caption{Predicted lattice thermal conductivity of the monolayers as a function of temperature with taking the isotope scattering into account.}
	\label{kappa}
\end{figure}

\begin{figure*}
	\centering{
		\includegraphics[width=0.85\textwidth]{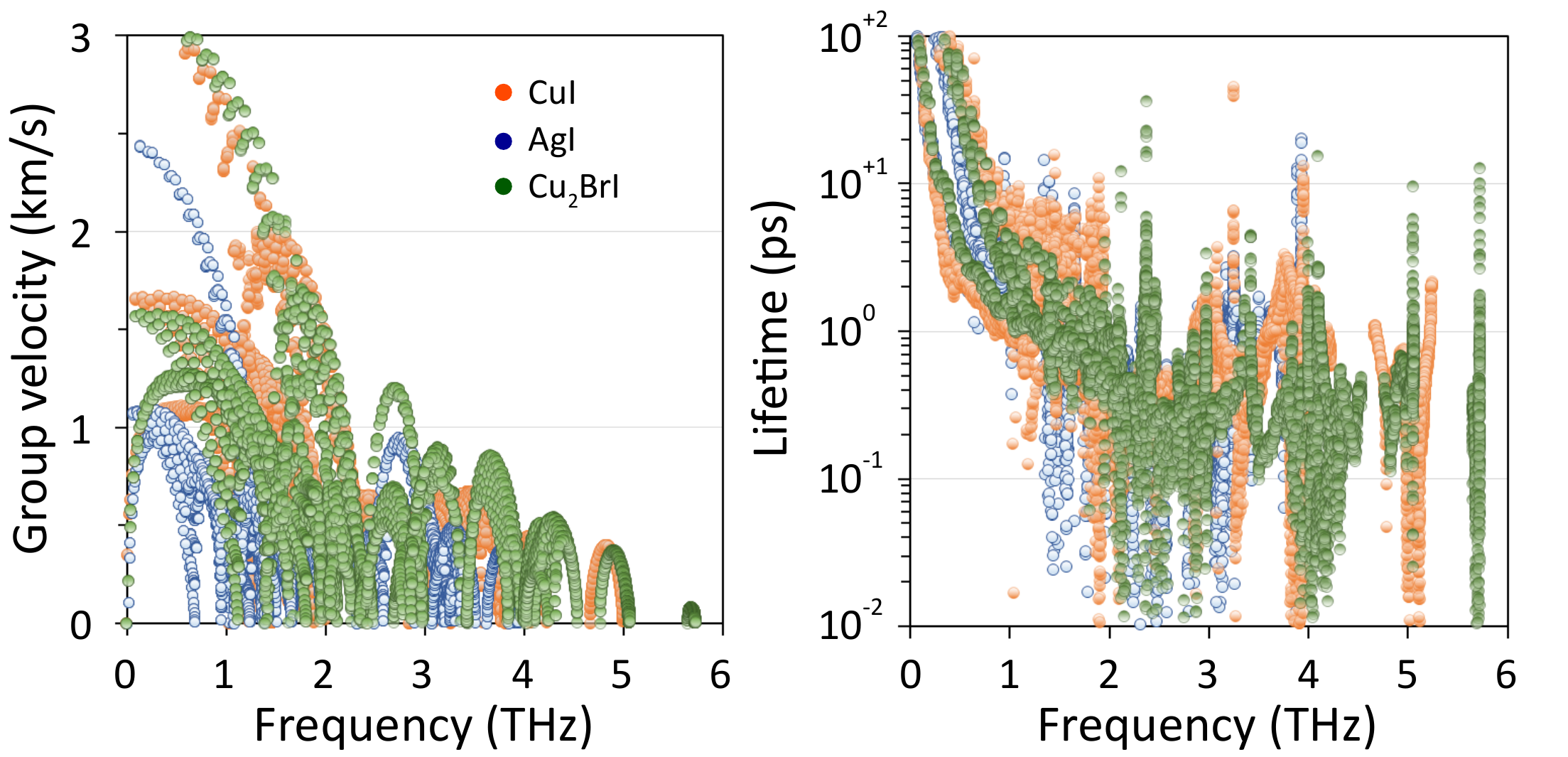}}
	\caption{Frequency dependent phonon group velocity (left panel) and lifetime (right panel) of CuI, AgI, and Cu$_2$BrI monolayers at \mbox{300 K}.}
	\label{time}
\end{figure*}

Owing to the dependence of the single-shot G$_0$W$_0$ method on the DFT \mbox{mean-field} starting points, we performed the fully \mbox{self-consistent} GW for CuI monolayer. As shown in Fig.~S7, updating the eigenvalues and eigenstates leads to a slight red shift in the imaginary part of the dielectric function. Also, the intensity of the peaks is reduced, however, the location and number of the peaks stay almost unchanged. As a result, the binding energy of the ground state exciton remains constant at $\sim$0.87 eV. Hence, we find that the obtained results are not dependent on the choice of input parameters.

The other optical coefficients of the monolayers i.e. the real part of the dielectric function, absorption coefficient, and reflectivity are calculated at the G$_0$W$_0+$BSE level. As shown in Fig.~S8 (a), the static dielectric constant of the monolayers is 1.36, 1.44, 1.29, and 1.32 for CuI, AgI, Cu$_2$BrI, and Ag$_2$BrI monolayers, respectively. That is to say, with increasing the average atomic number of the monolayer, the static dielectric constant increases. It is also seen that the real part of the dielectric function remains always positive in the entire range, showing the ultralow reflectivity of the monolayers. From the absorption coefficients, it is understood that the monolayers are good absorbers in the ultraviolet region \mbox{(3.26 to 10 eV)}. The mean value of absorption in this region is 1.37, 1.87, 1.16, and \mbox{1.44 $\times$ $10^7$ m$^{-1}$} for CuI, AgI, Cu$_2$BrI, and Ag$_2$BrI monolayers, respectively. Indeed, the structure with the largest (smallest) atomic number has the largest (smallest) absorption rate. From Fig.~S8 (c), it can be seen that the monolayers are highly transparent in the visible light region \mbox{(1.63 to 3.26 eV)}. In addition, the mean value of reflectivity in the entire region remains under 3, 4, 2, and 2\% for CuI, AgI, Cu$_2$BrI, and Ag$_2$BrI monolayers, respectively. Overall, for their band gaps, ultrahigh transparency, and flexibility, the monolayers are predicted to be very suitable for application as electron and hole transport layers in solar cell.

\subsection{Phononic properties}
We now investigate the temperature-dependent lattice thermal conductivity of CuI, AgI, Cu$_2$BrI, and Ag$_2$BrI monolayers, as it is illustrated in Fig.~\ref{kappa}. The results show isotropic lattice thermal conductivity in these novel monolayers, which is consistent with their highly symmetrical lattice. \mbox{At 300 K}, the thermal conductivity of CuI, AgI, Cu$_2$BrI, and Ag$_2$BrI monolayers taking the isotope scattering into account are predicted to be remarkably low as 3.75, 2.27, 3.13, and \mbox{1.26 W/mK}, respectively. We found that the lattice thermal conductivity in these systems follows the \mbox{$T^{-1}$} trend with respect to the temperature ($T$). The predicted values for thermal conductivity are also consistent with the classical theory saying a material with lower elastic modulus and higher atomic weight shows lower thermal conductivity. By increasing temperature up to \mbox{800 K}, due to the enhanced phonon scattering, the thermal conductivity of CuI, AgI, Cu$_2$BrI, and Ag$_2$BrI monolayers are reduced to 1.40, 0.85, 1.16, and 0.47 W/mK, which are very promising for thermoelectric applications.

To better understand the underlying mechanism resulting in the low lattice thermal conductivity of these novel 2D materials, in Fig.~\ref{time}, we compare their phonon group velocity and lifetime. As expected, with the weakening of the atomic bonds, the phonon group velocities are considerably suppressed in the AgI monolayer, which is more noticeable for low-frequency acoustic modes. This is consistent with the observed narrower dispersions for the acoustic phonon modes in AgI monolayer than those of CuI and Cu$_2$BrI counterparts. The phonon lifetime for these systems shows closer trends. It is noticeable that while the phonon group velocities are close for CuI and Cu$_2$BrI monolayers, the phonon lifetime is generally lower, particularly between \mbox{1$-$4 THz}, in Cu$_2$BrI monolayer than corresponding modes in the CuI counterpart, explaining its lower thermal conductivity.

\section{Conclusion}
Motivated by the successful synthesis of novel CuI and AgI monolayers via the graphene encapsulation approach,  herein, we carried out elaborated first-principles simulations to explore the key physical properties of non-Janus CuI and AgI and Janus Cu$_2$BrI and Ag$_2$BrI monolayers. We found that these novel 2D systems are stable, but are also soft materials with low elastic modulus. On the basis of full-iterative solution of the Boltzmann transport equation, the lattice thermal conductivity of CuI, AgI, Cu$_2$BrI, and Ag$_2$BrI monolayers at room temperature were predicted to be remarkably low, 3.75, 2.27, 3.13, and 1.26 W/mK, respectively, which are promising for thermoelectric applications. The electronic and optical properties were explored using many-body perturbation calculations. Particularly, the excitonic effects on the optical properties were taken into account by solving the Bethe-Salpeter equation. The optical gaps were obtained to be 3.14, 2.93, 2.64, and 2.62 eV for CuI, AgI, Cu2BrI, and Ag2BrI monolayers, which correspond to tightly bound excitons with binding energies of 0.88, 0.92, 0.95, and 0.99 eV, respectively. Effects of mechanical straining and electric filed on the tenability of the electronic and optical properties were also analyzed. The presented results provide an important and extensive vision for the key physical properties of the non-Janus CuI and AgI and the Janus Cu2BrI and Ag2BrI monolayers and may be valuable to the future applications of optoelectronic devices.

\section*{Acknowledgment}
B.M. and X.Z. appreciate the funding by the Deutsche Forschungsgemeinschaft (DFG, German Research Foundation) under Germany’s Excellence Strategy within the Cluster of Excellence PhoenixD (EXC 2122, Project ID 390833453). B.M. is greatly thankful to the VEGAS cluster at the Bauhaus University of Weimar for providing the computational resources.

\section*{Declaration of Interests}
The authors declare that they have no conflict of interest.
	
\bibliographystyle{apsrev4-2} % Tell bibtex which bibliography style to use
\bibliography{Ref}
\end{document}